\documentclass[prb,twocolumn,showpacs,preprintnumbers,amsmath,amssymb]{revtex4}
\usepackage{graphicx}
\usepackage{dcolumn}
\usepackage{bm}

\begin{document}

\preprint{cond-mat/}

\title{Dynamics of Density Imbalanced Bilayer Holes in the Quantum Hall Regime}

\author{S. Misra}
\author{N. C. Bishop}
\author{E. Tutuc}
 \altaffiliation{present address: Microelectronics Research Center, University of Texas, Austin}
\author{M. Shayegan}
\affiliation{
Department of Electrical Engineering, Princeton University, Princeton, NJ 08544
}

\date{\today}

\begin{abstract}
We report magnetotransport measurements on bilayer GaAs hole systems with 
unequal hole concentrations in the two layers. At magnetic fields where one layer is in the 
integer quantum Hall state and the other has bulk extended states at the Fermi energy, the 
longitudinal and Hall resistances of the latter are hysteretic, 
in agreement with previous measurements. For a fixed magnetic field inside this region and
at low temperatures ($T\le$ 350 mK), the time evolutions of the
longitudinal and Hall resistances show pronounced jumps followed by slow 
relaxations, with no end to the sequence of jumps. Our measurements demonstrate that 
the jumps occur simultaneously in pairs of contacts 170 $\mu$m apart, and appear to 
involve changes in the charge configuration of the bilayer. 
In addition, the jumps can occur with either random or regular periods,
excluding thermal fluctuations as a possible origin for the jumps. 
Finally, while remaining at a fixed field, we warm the sample to above 350 mK, where the jumps disappear.
Upon recooling the sample below this temperature, the jumps reappear, indicating that the jumps do not
result from nearly dissipationless eddy currents either. 
\end{abstract}

\maketitle

\section{Introduction}
The hallmarks of the integer quantum Hall effect, a quantized Hall resistance
and zero longitudinal resistance in two dimensional (2D) carrier systems at low temperatures,
are essentially independent of the sample characteristics. 
This universal behavior results from the Fermi
energy lying in between two Landau levels (LLs), meaning that the bulk has only 
localized states at the Fermi energy, and thus the sample's properties are dominated
by the extended one-dimensional edge states. \cite{PrangeSpringer1990} Accordingly, the
properties of the bulk have been difficult to access directly except in a series of recent 
scanning probe experiments. \cite{TessmerNature1998,IlaniNature2004,SteelePRL2005} 
These studies indicate that the bulk can be thought of as a set of isolated quantum 
dots and antidots seperated by incompressible regions.

A number of recent experiments have placed a probe, such as a single-electron 
transistor, \cite{HuelsPRB2004} a magnetometer, \cite{ElliotPRB2006,ElliotEPL2006} or even 
another conducting 2D layer, \cite{ZhuPRB2000,TutucPRB2003,PanPRB2005,SiddikiPhysica2006}
close to a 2D layer in the quantum Hall state (QHS) to explore how the 
isolated  bulk states achieve equilibrium. Here we focus on a bilayer 2D hole system 
with unequal hole densities such that, at a particular 
magnetic field, the Fermi energy lies in between LLs for one layer, and
near the middle of one LL for the other layer. In samples with narrow tunnel
barriers between the two layers
($w_b<$ 7.5 nm), the holes can easily tunnel between the two layers, and the system appears
to be in equilibrium. \cite{TutucPRB2003}
In contrast, for samples with wider barriers ($w_b>$ 7.5 nm), 
the magnetoresistance of the conducting layer, which we call the probe layer, 
is hysteretic for the range of magnetic fields where the other layer is in a QHS. 
\cite{ZhuPRB2000,TutucPRB2003,PanPRB2005,SiddikiPhysica2006}
This effect has been proposed to be the consequence of a non-equilibrium charge distribution.
\cite{ZhuPRB2000,TutucPRB2003,SiddikiPhysica2006}
The charge configurations of the QHS layer are different on the low- and high-field 
sides of the QHS, which result in different charge configurations for the probe layer as well. 
This charge configuration becomes frozen once the former enters the QHS, resulting in 
a hysteretic magnetoresistance. For bilayer hole samples with intermediate 
barrier widths (7.5 nm $< w_b <$ 200 nm), once the magnetic field is set such that 
one layer is in the QHS, the system appears to never reach equilibrium. Instead, 
both the longitudinal and Hall resistances of the probe layer show large jumps as a function of time:
the resistance changes by $\Delta r\sim50-500 \Omega$ 
over a short time (faster than 300 ms), typically followed by a slow 
relaxation ($\tau\sim40-400$ s), with no apparent end to the sequence of jumps. \cite{TutucPRB2003} 

The purpose of this paper is to present new data describing additional features of 
this non-equilibrium phenomenon. Section II covers the details of the experiment.
In Section III, we demonstrate that the resistance jumps occur simultaneously 
in pairs of contacts which are 170 $\mu$m apart, and appear to involve 
a change in the charge configuration of the bilayer. We show in Section IV that the jumps can occur at 
quasi-periodic time intervals, or, for slightly different experimental conditions, at
largely random time intervals. Data presented in Section V reveal that 
the jumps decrease in amplitude with increasing temperature, and disappear above $\sim$ 350 mK.
Surprisingly, when we recool the sample in a fixed magnetic field, the jumps reappear at low temperatures.
In Section VI, we conclude with a discussion of possible physical explanations of the data.

\section{Experimental Details}

We performed electrical transport measurements on double quantum well samples grown on GaAs
(311)A substrates. Although the behavior described here has been observed for a number
of hole bilayer samples having a range of barrier thicknesses and densities, \cite{TutucPRB2003}
we present data taken on three samples from one wafer. The wafer
contains a pair of 15 nm-wide GaAs quantum wells seperated by a $w_b =$ 11 nm AlAs
barrier, and flanked by a spacer and Si-doped layers of Al$_{0.21}$Ga$_{0.79}$As.
Sample A consists of a Hall bar with two current arms and 6 voltage probe arms, with an active region of
100 $\mu$m $\times$ 900 $\mu$m, as illustrated schematically in the inset to Fig. 1(a).
The distance between two adjacent contacts on one side is 170 $\mu$m,
and the width of the Hall bar is 100 $\mu$m. 
Samples B and C have simpler Hall bar configurations, with a pair of current contacts and 
two voltage contacts. Alloyed InZn contacts were used 
to contact both layers of the structure. Using a selective gate-depletion scheme, 
\cite{EisensteinAPL1990} we could also independently contact the bottom layer in Sample A, 
or either layer in Sample C. 

Electrical transport measurements were made at a temperature of 30 mK, unless otherwise
noted, using standard ac lockin techniques with a drive current of 1 nA at a frequency of 4.2 Hz. 
The as-grown densities for the three samples used in this manuscript 
are $8\times10^{10}$cm$^{-2}$ (top layer) and $6\times10^{10}$cm$^{-2}$ (bottom layer), 
and the typical mobility at 30 mK is  $\simeq 30$ m$^2/$Vs.
In order to determine the carrier densities of the two layers in Sample A, 
we compared the total density to the density of just the bottom layer, both extracted 
from the Hall resistance ($R_{xy}$) at low magnetic fields. In Sample B, we examined the Shubnikov-de Haas 
oscillations of the longitudinal resistance ($R_{xx}$) of the bilayer in order to determine the densities 
of the top and bottom layers. For sample C, we compared the Shubnikov-de Haas oscillations of $R_{xx}$
of the top and the bottom layers, acquired independently, to establish the carrier densities of the 
two layers. The density of the bottom layer was then 
set such that it had a filling factor $\nu$ in between 1 and 2 at a 
magnetic field where the top layer had a filling factor $\nu$ = 1. Thus, the Fermi energy of the bottom layer 
is in the second LL for a range of fields where the Fermi energy of the top layer is in between the first 
and second LLs. Unless noted otherwise, we report on the 
resistance of just the bottom layer, which we refer to as the probe layer, while the other layer, which we 
refer to as the QHS layer, is held at ground using a common ground contact. For Sample B, we show the 
resistance of the bilayer, which is indicative of the resistance of the probe layer. 
At magnetic fields where the QHS layer is near $\nu$=1, current flowing through
this layer causes no drop in longitudinal voltage. As a result, the voltage drop sensed across two
{\it bilayer} longitudinal contacts occurs solely in the probe layer.

\section{Layer Charge Instability}

Figure 1(a) shows the magnetoresistance of the 
probe layer of Sample A at 30mK, when the {\it probe} layer has a density $p_p = 7.4 \times 10^{10}$ cm$^{-2}$
and the other ({\it QHS}) layer has a density $p_q = 5.8 \times 10^{10}$ cm$^{-2}$.
As seen in this figure, the probe layer exhibits the zero longitudinal resistance and quantized Hall 
resistance characteristic of the integer quantum Hall effect. In addition, both the longitudinal and Hall
resistances of the probe layer are seen to be hysteretic in the range of magnetic fields where
the other layer is in the $\nu=1$ QHS. The hysteresis is
consistent with earlier reports from a single-well sample with an
unintentional parasitic layer, \cite{ZhuPRB2000} and recent measurements of intentionally imbalanced bilayer hole
\cite{TutucPRB2003} and electron \cite{PanPRB2005,SiddikiPhysica2006} samples. 

\begin{figure}
\includegraphics{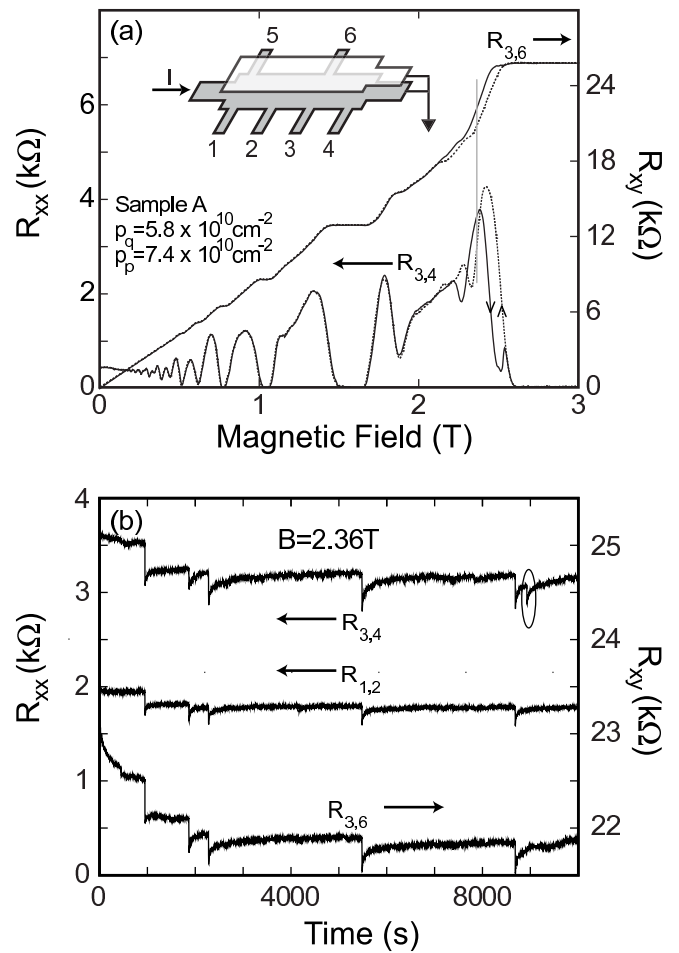}
\caption{ For Sample A, we adopt the convention that $R_{i,j}$ refers to the resistance measured between 
contacts $i$ and $j$ of the Hall bar (inset).
(a) The longitudinal ($R_{3,4}$) and Hall ($R_{3,6}$) resistances 
of the probe layer are hysteretic for a range of fields where the other layer is in the QHS. 
The densities of the QHS and probe layers are given by $p_q$ and $p_p$, respectively. 
(b) The time evolution of the longitudinal resistances $R_{1,2}$ and $R_{3,4}$,
and the Hall resistance $R_{3,6}$ taken simultaneously after sweeping the magnetic field up 
to 2.36 T, indicated by a vertical line in (a). 
The circled jump is the only one which does not occur simultaneously 
in all contact pairs.}
\end{figure}

This hysteresis is believed to result from a non-equilibrium interlayer charge distribution. 
\cite{ZhuPRB2000,TutucPRB2003,SiddikiPhysica2006}
This can be understood by examining how the Fermi energy of two layers changes 
as a function of the magnetic field, as illustrated schematically in Fig. 2. Because the probe layer has a 
larger density than the QHS layer at zero magnetic field, we
show the bottom of the subbands for the two wells as being offset from one another, but the Fermi energies
as being equal (at zero magnetic field). Upon increasing the magnetic field
just below the point where the QHS layer is at filling factor $\nu = 1$, the Fermi energy of this layer
has increased more than that of the probe layer. This results in a redistribution of charge between the
two layers, such that the QHS layer will have a higher carrier density
relative to what it had at zero magnetic field, and the probe layer a lower density. 
By the same arguement, upon {\it decreasing}
the magnetic field to just above the point where the QHS layer is at $\nu = 1$, 
its Fermi energy has decreased relative to what it was at zero magnetic field. Thus, the QHS 
layer will have a lower carrier density relative to what it had at zero magnetic field, and the probe layer a higher density.
Upon entering the QHS, the charge configuration 
of the QHS layer becomes frozen, as there are now only localized states at the Fermi 
energy, and large incompressible regions seperate these states from a reservoir. This layer exerts an 
electrostatic potential on the probe layer, and thus the
density of the latter becomes trapped at a non-equilibrium level as well. 
Because its density is lower (higher) than at zero magnetic field when sweeping the magnetic field up 
(down), the magnetoresistance of the probe layer should look shifted to lower (higher) fields,
consistent with the data shown in Fig. 1(a). 
At even higher magnetic fields, the Fermi energy of the QHS layer will remain in the first LL, while the
probe layer will enter the $\nu = 1$ QHS, that is, their roles reverse. Just below the field where
the probe layer enters the $\nu = 1$ QHS, it has a higher Fermi energy than the QHS layer. Just above
the field where it enters the $\nu = 1$ QHS, it has a lower Fermi energy than the QHS layer. This
situation is thus identical to the one described above, and thus we would expect the magnetoresistance
of the QHS layer to look shifted to lower (higher) fields when sweeping the magnetic field up (down). 
Note in particular that this picture is also consistent with the data presented in Ref. \onlinecite{TutucPRB2003}.

\begin{figure}
\includegraphics{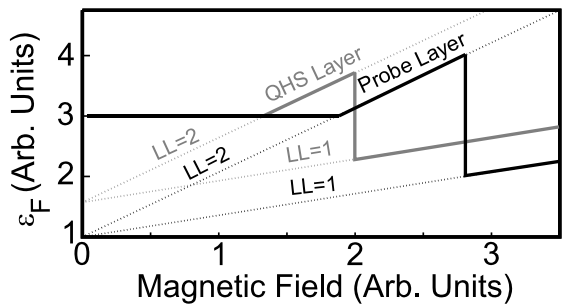}
\caption{Schematic diagram of the change of the Fermi energy as a function of magnetic field
for the probe layer (black) and the QHS layer (gray), referenced to the Fermi energy of the 
respective 
layer at zero magnetic field. Also shown as dotted lines are the energies of the first and second
LLs for both layers.}
\end{figure}

In agreement with the results reported in Ref. \onlinecite{TutucPRB2003} for
bilayer hole samples with intermediate 
barrier widths (7.5nm $< w_b <$ 200nm), we find that the bilayer system shows peculiar 
dynamics in the hysteretic region of magnetic fields. As shown in Fig. 1(b), for example, the time 
evolution of the resistance after stopping a field sweep in the hysteretic region
features sudden jumps in resistance, where the resistance changes by 
$\Delta R \approx \pm 25$ to $\pm500 \Omega$ over a short time scale (as fast as 300 ms) 
that is limited by our experimental bandwidth. A jump is followed by a slow 
relaxation over a long time scale ($\tau\approx 20-400$ s), and then another jump 
a time ($\Delta t \approx 30-3000$ s) 
later. This sequence of jumps continues for as long as we have tracked the time evolution, 
up to $\sim10^5$ s, with no systematic change in the jump amplitude or sign, the time spacing 
between jumps, or the relaxation time constant. 
Consistent with previous results, \cite {TutucPRB2003} we find that these characteristics 
depend sensitively on the magnetic field and carrier concentration, 
and are different for separate cooldowns. The data indicate that the dynamics of the sample 
itself are responsible for the jumps. No such jumps are seen in a resistor 
hooked up in series with the sample, or when measuring the sample at a 
magnetic field outside the hysteretic region. 
None of the characteristics of the jumps change significantly when varying the drive current 
used to measure the resistance (0.5 nA 
- 8nA), up to the point where the drive current starts to heat the sample. 
Most strikingly, the jumps happen whether or not we probe the resistance, as shown in 
Fig. 3. Even when grounding the sample
for hundreds of seconds, and then resuming our measurement, we can see the decay associated 
with a jump which must have happened while the sample was grounded.

\begin{figure}
\includegraphics{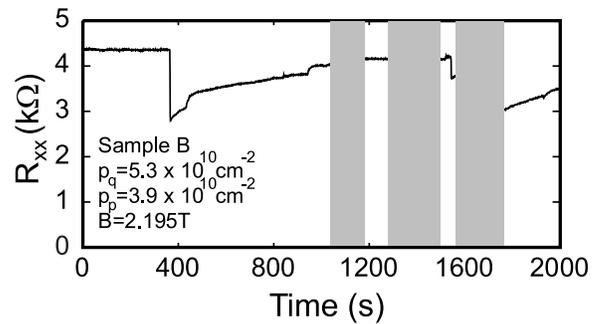}
\caption{ The time evolution of the longitudinal resistance of the bilayer, where the contacts
to the sample are grounded for the time periods shown in gray. Note in particular that
a jump must have occured during the third time period, even though the sample was grounded. }
\end{figure}

The jumps themselves signal a significant change in the properties of the probe layer.
In Fig. 4, we compare magnetoresistance traces taken from Sample B
when sweeping the magnetic field at two different sweep rates, one fast enough that no jumps
occur while sweeping the field, and a second slow enough that many jumps occur. 
The fast magnetoresistance traces (dotted lines), which contain no jumps, 
define the two branches of the hysteresis loop, the upward and downward branches. 
When sweeping the magnetic field slowly enough so that the resistance does jump 
during the field sweep (solid line), we find that the magnetoresistance trace departs from the
branch of the hysteresis loop on which it would be expected to lie. For the data in Fig. 4, when
we slowly sweep the magnetic field up, we find that the jumps first cause this 
magnetoresistance trace to depart from the upward branch of the hysteresis loop, 
to a point where the magnetoresistance is not even at a value in between the extremes 
defined by the upward and downward branches of the hysteresis loop.
Later, further jumps result in the magnetoresistance joining the {\it downward} branch 
of the hysteresis loop despite being taken while sweeping the magnetic field {\it upwards}. 
Although it is more common that jumps shift the magnetoresistance to curves unrelated to the 
upward or downward branch of the hysteresis loop, this data set clearly demonstrates 
that the jumps have changed the carrier concentration of the probe layer. 
Within the layer charge instability scenario, the data would suggest that there are a number of 
bilayer charge configurations which produce different
magnetoresistance curves. These magnetoresistance curves
have similar features, but appear shifted in magnetic field, suggesting that these charge
configurations involve different carrier concentrations for the probe layer. The jumps would then
result from sudden changes in the probe layer charge configuration. 

\begin{figure}
\includegraphics{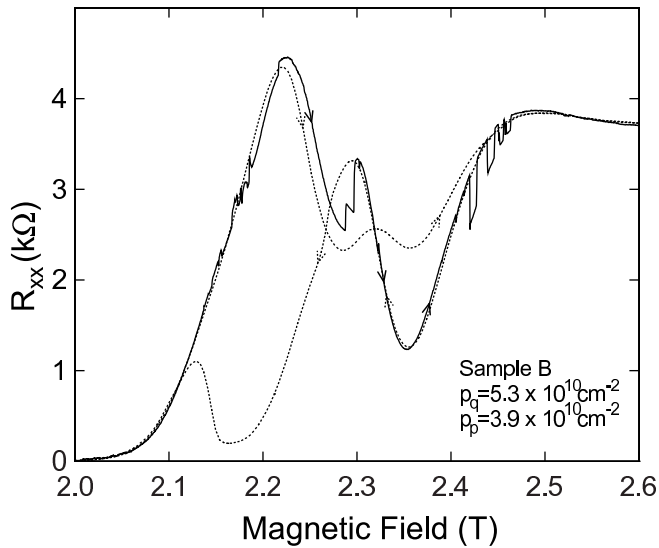}
\caption{Solid trace: the longitudinal magnetoresistance of the bilayer taken while sweeping the
magnetic field up at a rate of 0.17 mT/s. Dotted traces: 
magnetoresistance of the bilayer while sweeping the field at a rate of 3.3 mT/s in the
direction indicated by the arrows.}
\end{figure}

We next address whether these jumps are local fluctuations of the charge configuration by 
measuring the resistance simultaneously in three sets of contacts. 
As shown in Fig. 1(b), we find that the jumps almost always occur at 
the same time in all three sets of contacts, which are up to 170 $\mu$m apart. 
This indicates that the jumps are not the result of a local fluctuation, and implies that
there is a change in either the QHS layer or the probe layer over large length scales which
is responsible for creating a jump in the probe layer resistance.
However, the direction and amplitude of the jumps, and the 
relaxation time constant, are not, in general, the same for the jumps seen simultaneously 
in different contacts. We also occasionally observe a jump in only one of 
the longitudinal sets of contacts, or in the Hall contacts, 
without any jump in the other two sets of contacts. These observations suggest that the 
charge configuration is changing over a length scale which, while large,
does not extend across the entire sample.

\section{Quasi-Periodic Oscillations}

\begin{figure*}
\includegraphics{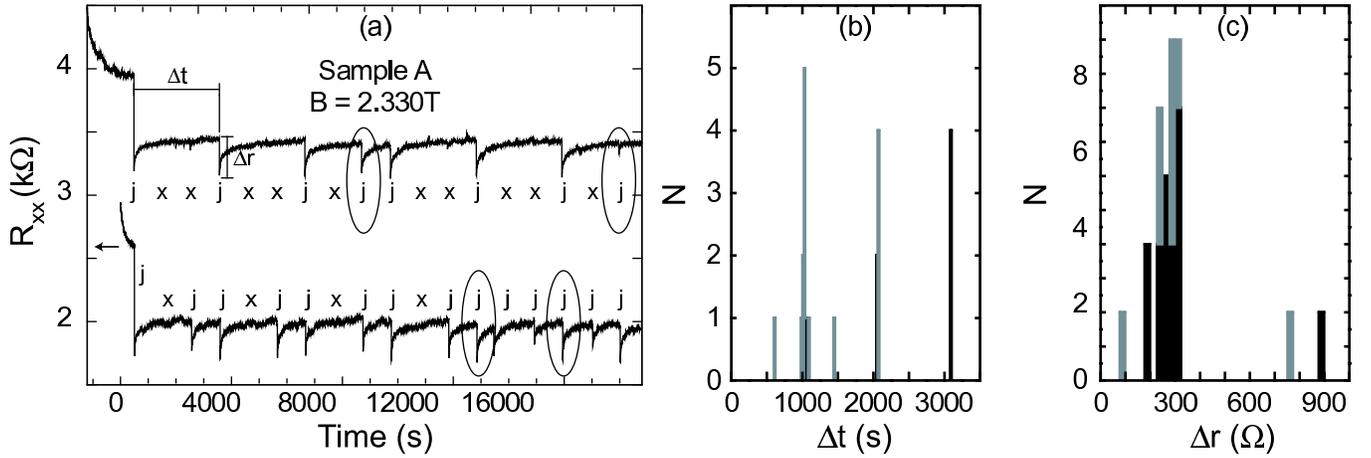}
\caption{ (a) Time evolution of $R_{1,3}$ for the probe layer of Sample A
taken immediately after ramping the magnetic field from 0 T up to 2.33 T. The layer densities  are
$p_p=7.4 \times 10^{10}$cm$^{-2}$ and $p_q=5.8 \times 10^{10}$cm$^{-2}$. 
The experiment was repeated twice, and the two sets of time evolution data 
have been offset vertically for clarity, and horizontally to match the position of the first jump. 
(b) Stacked histogram of the time $\Delta t$ between consecutive jumps in the time captures shown in (a), 
with black (gray) bars corresponding to $\Delta t$ extracted from the top (bottom) trace. 
Taking the basic time block to be 1030 seconds, the train of jumps in the top trace of part (a) follows the 
sequence jump (j), no jump (x), no jump (x), with two 
jumps occuring out of sequence (circled). 
The train of jumps for the bottom time capture shown in part (a) 
follows the sequence j-x-j, 
with two extra jumps occurring out of sequence (circled). The first of the circled jumps in the lower trace of (a)
is the only one whose $\Delta t$ is not an integer multiple of 1030 s.
(c) Stacked histogram of the size of a resistance jump $\Delta r$. Excluding the jump that starts 
the sequence, the standard deviation of jumps sizes (55 $\Omega$) is much smaller than the median 
jump size (285 $\Omega$).}
\end{figure*}

Thermal fluctuations are the 
most obvious candidates for what drives the imbalanced bilayer to jump between various charge
configurations. The physical situation where a system has two nearly energy degenerate charge 
configurations in a quantizing magnetic field has been studied previously in resonant transport through a 
quantum dot. \cite{KouwenhoevenPRL1994, KouwenhoevenPRB1997} In these studies, the quantum dot has two 
charge configurations with similar energies, one which has a high conductance and the other a low 
conductance. The fluctuations in the charge configuration of the quantum dot result in a conductance 
through the dot which switches between the high conductance and low conductance values to create telegraph 
noise. As is to be expected for a process driven by fluctuations, 
the amount of time between switching events is rather widely distributed. 

\begin{figure*}
\includegraphics{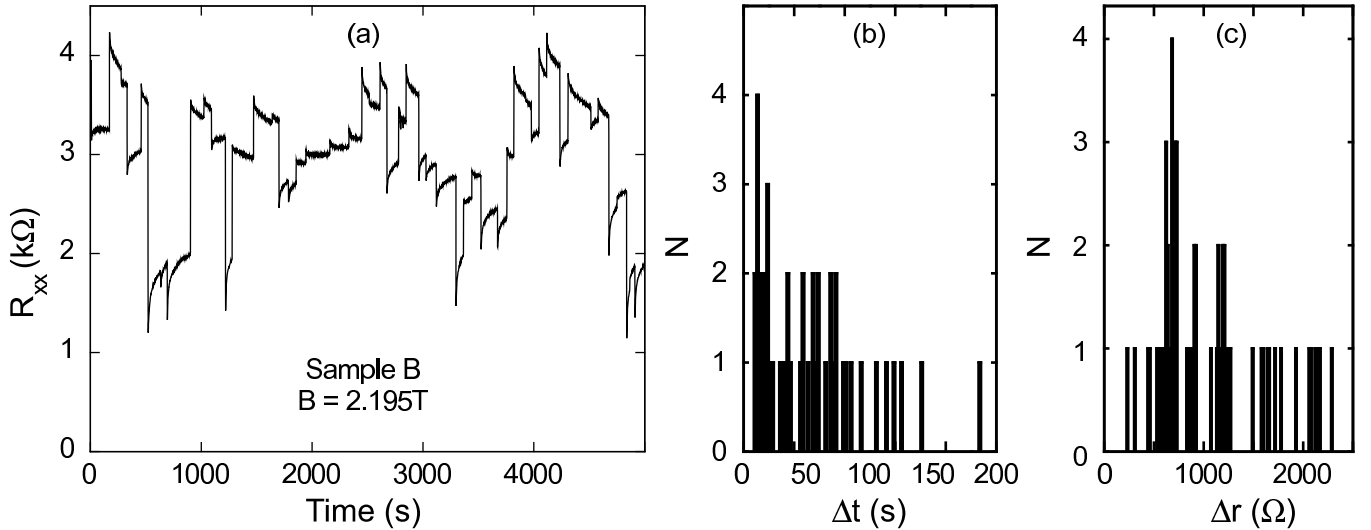}
\caption{ (a) The time evolution of Sample B, having layer densities 
were $p_p=5.3 \times 10^{10}$cm$^{-2}$ and $p_q=3.9 \times 10^{10}$cm$^{-2}$. 
(b) Histogram of the time $\Delta t$ between 
consecutive jumps in the time capture shown in (a). (c) Histogram of the size of a jump $\Delta r$.
The standard deviation of jump sizes (500 $\Omega$) is nearly as large as the median (575 $\Omega$). }
\end{figure*}

In order to determine whether fluctuations are responsible for the jumps in resistance seen in our 
imbalanced bilayer system, we examine the distribution of time 
elapsed in between jumps.  In Fig. 5(a), we 
show the time evolution of the resistance taken immediately after sweeping the magnetic field into the 
hysteretic range of fields and stopping at a particular field. Each of the two time evolutions shown 
in Fig. 5(a) were recorded after ramping the magnetic field from 0 T to 2.33T, and then recording
$R_{xx}$ as a function of time. As shown in Fig. 5(b), the time elapsed 
between successive jumps, $\Delta t$, is not widely distributed, but rather is almost always either 1030 
$\pm$ 40 s, 2060 $\pm$ 10 s or 3090 $\pm$ 5 s. Reexamining Fig. 5(a) data
in blocks of 1030 s periods, the top trace shows a jump (j), followed by no jump (x), and then again
by no jump (x). This j-x-x pattern repeats every 3090s. The bottom trace shows a j-x-j pattern
which repeats every 3090s. In all cases, this pattern
becomes less robust the longer we take the data, although the quasi-periodicity remains. 
Comparing the two traces and shifting them in time so that their first jump lines up, as shown in Fig. 
5(a), reveals that all 8 jumps seen in the top trace occur within 20 s of a 
jump in the bottom trace. Combined, these observations indicate the presence of a reliable time scale for 
the jumps, and thus exclude the possibility that thermal fluctuations
are driving the system to switch between a set of nearly energy degenerate charge configurations. The
distribution of resistance jump amplitudes, shown in Fig. 5(c), is relatively sharp, inasmuch as the standard 
deviation of jump sizes (55 $\Omega$) is much smaller than the median jump size (285 $\Omega$).

The quasiperiodic behavior of jumps seen in Fig. 5, however, is not always observed. 
In the time evolution data for Sample B, \cite{TutucPRB2003} for example, shown in Fig. 6(a), 
the system appears to be jumping between a significantly larger number of quasi-stable points, as jumps 
occur in both directions, and do not appear to relax to the same value. The histogram of the time between 
jumps, shown in Fig. 6(b), is not sharply peaked at a small number of values, as in Fig. 5(b), but is 
rather widely distributed. Quantitatively, the distribution of jump 
amplitudes for the data in Fig. 6 is wide, as can be seen by comparing the standard deviation of jump amplitudes with the mean 
jump amplitude. 
Finally, we note that the occurrance of a semi-regular period and a narrow 
distribution of jump sizes, as in Fig. 5,
as opposed to a largely random period and wide distribution of jump sizes, as in Fig. 6, 
are not mutually exclusive. Curiously, even when the system has a wide range of jump 
amplitudes, it can 
exhibit quasi-periodic time evolution. In Figs. 7 and 8, which will be discussed in Section V, the system 
appears to be jumping between a number of quasi-stable points; the jumps relax to different values of 
resistance. Similar to the data shown in Fig. 5(b), and unlike the data in Fig. 6(b), 
the distribution of time between successive jumps shows
quasi-periodic behavior. However, unlike the data in Fig. 5(c), and similar to the data in Fig. 6(c), the
distribution of jump amplitudes is broad.

\section{Temperature Dependence}

\begin{figure}
\includegraphics{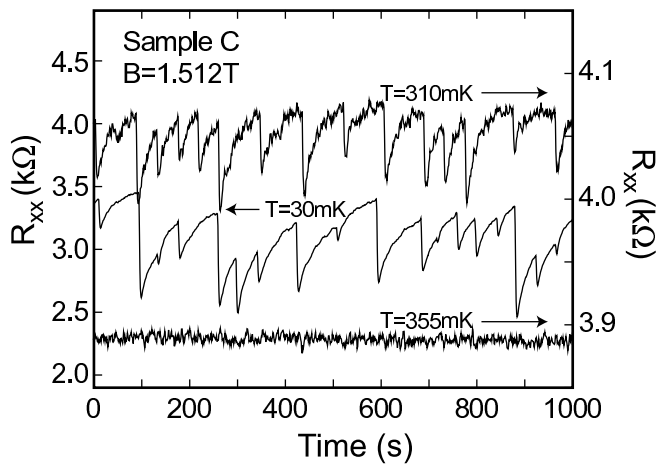}
\caption{ Time evolution of the longitudinal resistance of the probe layer at 1.512 T in Sample C,
with layer densities of $p_p=6.7 \times 10^{10}$cm$^{-2}$ and $p_q=4.1 \times 10^{10}$cm$^{-2}$. 
Shown are representative 1000 s slices of 5000 s-long sweeps taken at 30 mK, 310 mK and 355 mK.}
\end{figure}

\begin{figure}
\includegraphics{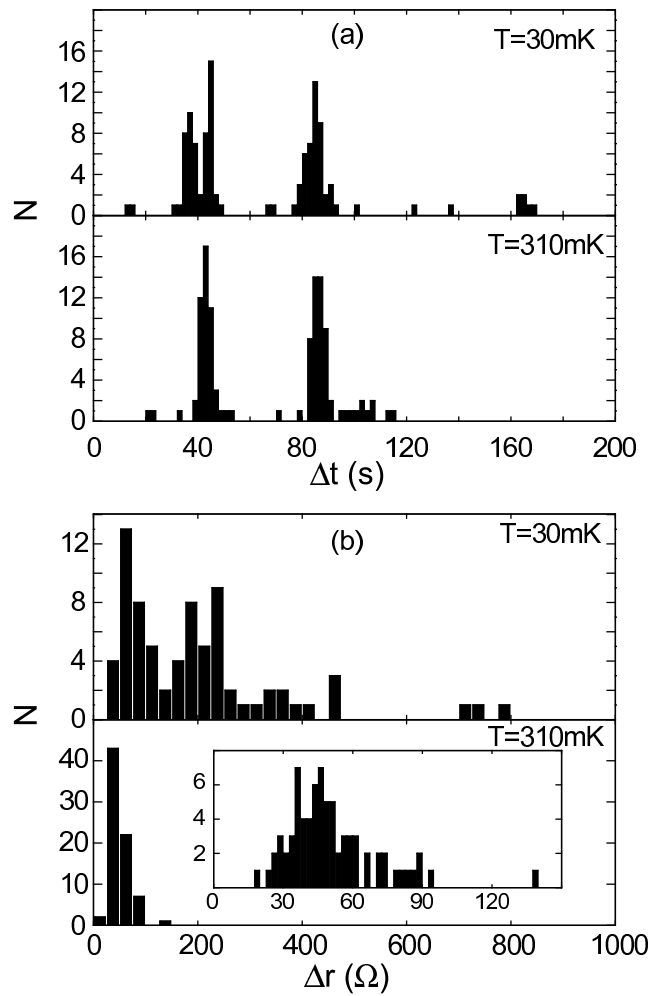}
\caption{Statistics for the jumps seen in the time evolution data shown in Fig. 7.
(a) Histograms of the time $\Delta t$ between consecutive jumps.
(b) Histograms of the jump size $\Delta r$. 
The inset shows a magnified histogram of the 310 mK data. Note that we
use a bin size of 25 $\Omega$ in the main part of (b), while we use a bin size of 
2.5 $\Omega$ in the inset. Note also that all the histograms shown in this figure are for the entire, 
5000 s-long, traces whose 1000 s slices are shown in Fig. 7.}
\end{figure}

The non-equilibrium behavior of the probe layer is intimitely tied to the other layer in the bilayer
being in the $\nu = 1$ QHS. Because the strength of the QHS decreases with increasing temperature, 
we expect the properties of the jumps to change upon increasing the temperature of the system. 
In Figs. 7 and 8, we examine how the jumps change upon increasing the temperature for Sample C. 
The magnetic field was first swept up to 1.512 T, and five thousand seconds 
elapsed before taking the first time evolution at $T=$ 30 mK (center trace in Fig. 7). 
While remaining at this field, we then warmed 
the sample up slowly, tracking the time evolution at a number of temperatures. We find that 
the jumps decrease in amplitude upon increasing the temperature, but otherwise are 
qualitatively similar to how they appear at 30 mK. In Fig. 7 (upper trace),
we show the time evolution at  310 mK, where the jumps are clearly visible, 
although their amplitude is significantly smaller than at 30 mK. As can be seen by examining
the histogram of time between jumps (Fig. 8(a)), the quasiperiodic nature of the jumps
at this field is unaffected by raising the temperature. 
What does change dramatically is the amplitude of the jumps, as shown in Fig. 8(b), which drops from 
hundreds of Ohms at 30 mK to tens of Ohms at 310 mK.  The time constant for the slow relaxation 
after a jump, determined by fitting the relaxation to a double exponential, as in Ref. \onlinecite{TutucPRB2003}, 
remains roughly 40 s at both temperatures. However, the temperature independence of the relaxation time
seen here is not a robust feature of the data. For Sample A, at a magnetic field of 1.87 T and layer 
densities of 5.1 and  7.4 $\times 10^{10}$cm$^{-2}$, the relaxation time was seen to vary between 400 s at 
a temperature of 30 mK and 10 s at 270 mK. 

\begin{figure}
\includegraphics{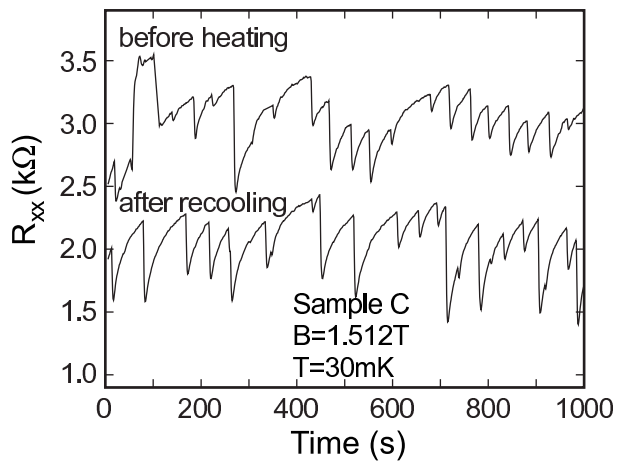}
\caption { Representative time evolution of the longitudinal resistance of the probe layer at 1.512T in Sample C.
The top trace, which is a different 1000 s slice of the same 5000 s sweep used in Fig. 7, 
was first recorded at 30mK. Then, with the magnetic field fixed at 1.512T, 
the sample was heated to a temperature (355mK) where the jumps disappear. We then recooled the sample back 
to 30mK, and recorded the time evolution (1000 s of which is shown in the bottom trace) 
after reaching base temperature again.}
\end{figure}

\begin{figure}
\includegraphics{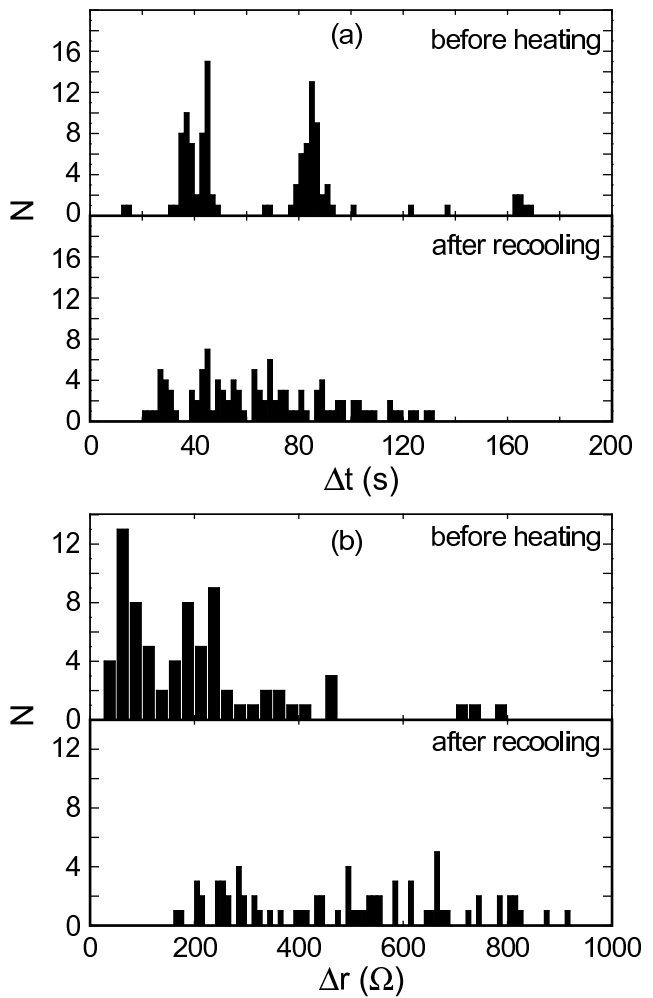}
\caption { Statistics for the jumps seen in the time evolution data shown in Fig. 9. 
(a) Histograms of the time $\Delta t$ between consecutive jumps.
(b) Histograms of the jump size $\Delta r$. Note that the same trace was analyzed for the
"before heating" condition as was analyzed for the 30 mK data in Fig. 8. 
}
\end{figure}

At high enough temperatures, the QHS layer begins to weakly conduct.
At that point, any non-equilibrium condition induced
by the field sweep should dissipate.
We heated the sample above the temperature (355 mK, lower trace on Fig. 7 (a)) where the jumps and the hysteresis
in the probe layer magnetoresistence disappear, allowing the sample 
to come to equilibrium. In Figs. 9 and 10,
we compare two sets of data taken at a fixed magnetic field, the first set before heating the sample, and the second
after heating the sample to 355 mK and then cooling back to 30mK over three hours.
Surprisingly, when we recool the sample, 
the jumps in the time evolution resume. But the histogram of times in between jumps has also changed from 
being quasi-periodic to being nonperiodic after recooling.

\section{Summary and Discussion}

We have examined the magnetoresistance of bilayer hole systems whose layer densities have
been intentionally set to be unequal. At magnetic fields where one layer is in the $\nu = 1$ QHS
and the other layer (the probe layer) has bulk extended states at the Fermi energy, the longitudinal and Hall 
resistances of the latter are hysteretic, in agreement with previous studies. 
\cite{ZhuPRB2000,TutucPRB2003,PanPRB2005,SiddikiPhysica2006} For a fixed magnetic field inside 
this hysteretic 
region, and at low temperatures, the resistance of the probe layer shows pronounced jumps followed by a slow 
relaxation, with no end to the sequence of jumps, also in agreement with previous work. \cite{TutucPRB2003}
The data presented in Section III suggest that the probe layer resistance jumps in response to the system
switching between different charge configurations. 
In addition, we have shown that the resistance jumps almost always 
occur simultaneously in pairs of contacts which are 170 $\mu$m apart, suggesting that the charge configuration of 
a significant part of the probe layer changes. 

In Section IV, we have demonstrated that the jumps can occur at
quasi-periodic time intervals, or, for slightly different conditions, at largely random time intervals.
The observation of quasi-periodic jumps excludes the possibility that the jumps represent thermal fluctuations 
between a set of nearly energy degenerate charge configurations, as such jumps would be expected to occur
at random time intervals. The presence of a reproducible time scale for a cycle that includes a jump 
followed by a relaxation is a defining characteristic of relaxation oscillators. In a typical
relaxation oscillator, a voltage builds up across a capacitor until a breakdown threshold is exceeded,
at which point the capacitor discharges. Such behavior has been seen in a circuit containing
a capacitor in parallel with a Corbino disk in the QHS. \cite{NachtweiPRB2003}
In this system, the capacitor charges until the QHS breaks down, which leads to 
a discharging of the capacitor until the QHS in the Corbino disk can be established again. Relaxation
oscillations have also been seen in the breakdown of the reentrant integer QHSs, 
\cite{CooperPRL2003} and in the transition from a pinned to a sliding Wigner solid phase. \cite{GaborPRL2007}
The presence of nonperiodic jumps is difficult to understand in terms of a single relaxation
oscillator, which should always have a well-defined period. One possibility is that our system contains
a number of relaxation oscillators in different parts of the sample. The charge-discharge cycles for 
different parts of the sample could be interdependent, masking any periodic behavior. Such behavior has been
seen before in a circuit consisting of two coupled, ac driven relaxation oscillators. \cite{GollubScience1978}
In addition to periodic behavior associated with the charge-discharge cycles of each relaxation oscillator, 
Gollub and coworkers \cite{GollubScience1978} 
found period-multiplying behavior, similar to the j-x-j and j-x-x patterns we see
in Fig. 5, and nonperiodic behavior for different ac signals applied to their circuit.

In Section V, we have shown that, while remaining at a fixed magnetic field, the jumps decrease in 
magnitude when increasing the temperature, disappearing for $T>$ 350 mK. Surprisingly, upon recooling the 
sample at fixed magnetic field, the jumps reappeared at low temperatures. 
This excludes any interpretation that relies on sweeping the magnetic field to establish a non-equilibrium
initial condition as a source for the jumps.
For example, a series of recent experiments have found that sweeping the magnetic field
sets up eddy currents in a 2D layer which can persist for {\it days} when it is in the QHS. 
\cite{HuelsPRB2004,ElliotPRB2006,ElliotEPL2006}
In one of these experiments, \cite{HuelsPRB2004} a single electron transistor placed close to a single 2D 
electron layer was used to show that there are sudden jumps in the local 
Fermi energy associated with these nearly dissipationless eddy currents breaking down the QHS
in the 2D layer. Such a breakdown of the QHS, driven by nearly dissipationless eddy currents, could lead to
jumps in the probe layer resistance seen by us.
However, for our experiment, by warming our sample to a 
temperature where neither the jumps nor the hysteretic magnetoresistance are seen, while keeping the magnetic
field fixed, we would have allowed the eddy currents to dissipate. Thus, the reappearance of jumps upon 
recooling the sample in fixed fields excludes the possibility that eddy currents
are responsible for the jumps we see in the probe layer resistance. 

The data presented here argue for a different physical origin for the jumps seen in our bilayer hole samples. 
Our data show that thermal fluctuations are not responsible for the jumps, and it is unlikely that quantum
fluctuations are either.
Another possibility is that two different energetic requirements are competing against one another to
create the jump-relaxation cycle. The 
data in Ref. \onlinecite{TutucPRB2003} show that the time the system spends in the relaxation part of the cycle 
increases when increasing the width of the barrier between the two layers, suggesting that the inter-layer
Coulomb interaction is involved in this part of the cycle. Because the amplitude of the
jumps decreases with increasing temperature, along with the strength of the $\nu = 1$ QHS, it seems likely that
the jump part of the cycle occurs within the QHS layer. The intra-layer Coulomb interaction thus likely plays an 
important role in the jump part of the cycle. 
We propose that the dynamic competition between the inter- and intra-layer
Coulomb interaction in the localized states of the QHS layer creates a charge-discharge cycle, which manifests 
itself as the relaxation-jump cycle. 
The capacitive coupling between the probe layer and the localized
states of the QHS layer could lead to a charging of the localized states. Thus, the inter-layer Coulomb 
interaction is responsible for charging. Eventually, the localized 
states in one region of the QHS layer become overcharged compared to neighboring regions, and discharge to them, 
or, in an avalanche, all the way to the edge state. The intra-layer Coulomb interaction is thus 
responsible for the discharge. The data shown here is 
consistent with the characteristics of such a charge-discharge cycle.
This cycle involves a change in the charge distribution of the QHS layer, 
and, through the inter-layer Coulomb interaction, the probe layer as well.
The sample itself could contain several such regions, each of which could cycle independently, 
yielding periodic time evolution, or they could be coupled, yielding complex time evolutions. 
Finally, at elevated temperatures, the thermal energy would lower the threshold where one region could 
discharge to neighboring regions, decreasing the amount of excess charge that could build up in any given
region. 

Ultimately, the physical processes responsible for the jump and the relaxation remain unclear. 
We emphasize that what {\it is} clear is that the 
bulk states of the QHS do not come to equilibrium over extremely long time scales in a
wide range of samples. This has been demonstrated clearly by 
Huels {\it et al.} \cite{HuelsPRB2004} in single layer 2D electron samples, and by us here in bilayer hole samples
with barriers larger than 7.5 nm. More broadly, it is possible that the bulk states of the QHS are out of 
equilibrium in general, but that very few experiments are sensitive to the non-equilibrium character of the 
bulk states. We add that we have been able to identify
cases where our bilayer samples do not show any indication of non-equilibrium
behavior: in imbalanced bilayer hole samples with barriers smaller than 7.5 nm, we observe no magnetoresistance 
hysteresis or resistance jumps. This implies that the bulk states of the QHS can come to 
equilibrium with the probe layer if interlayer tunneling is large enough.

We thank the DOE and the Princeton NSF MRSEC for support, and D. A. Huse and A. H. MacDonald for illuminating discussions.

\bibliography{misra}

\end{document}